\documentstyle[prl,aps,multicol,subfigure,psfig,subeqnar]{revtex}

\newcommand{\cL} { {\cal L} }
\newcommand{\dn} {\hskip 0.01in {\rm dn} \hskip 0.01in}
\newcommand{\cn} {\hskip 0.01in {\rm cn} \hskip 0.01in}
\newcommand{\sn} {\hskip 0.01in {\rm sn} \hskip 0.01in}

\newcommand{\demi}{\frac{1}{2}}

\newcommand{\be}{\begin{equation}}
\newcommand{\ee}{\end{equation}}
\newcommand{\ba}[1] {\begin{array}{ #1 }}
\newcommand{\ea}{\end{array}}

\begin{document}

\title{Stability of Repulsive Bose-Einstein Condensates in a Periodic Potential}
\author{J. C. Bronski$^{1}$, L. D. Carr$^{2}$,
B. Deconinck$^{3}$, J. N. Kutz$^{3}$\cite{byline}, and K. Promislow$^{4}$ \\}
\address{$^{1}$Department of Mathematics, University of Illinois 
         Urbana-Champaign, Urbana, IL 61801, USA\\}
\address{$^{2}$Department of Physics, University of Washington, 
         Seattle, WA 98195-1560, USA\\}
\address{$^{3}$Department of Applied Mathematics, University of 
         Washington, Seattle, WA 98195-2420, USA\\}
\address{$^{4}$Department of Mathematics, Simon Fraser University,
         Burnaby, B.C., CANADA V5A 1S6\\}
\maketitle

\date{\today}

\begin{abstract}
The cubic nonlinear Schr\"odinger equation with repulsive nonlinearity and
an elliptic function potential models a quasi-one-dimensional repulsive
dilute gas Bose--Einstein condensate trapped in a standing light wave.  New
families of stationary solutions are presented.  Some of these solutions have
neither an analog in the linear Schr\"odinger equation nor in the
integrable nonlinear Schr\"odinger equation.  Their stability is
examined using analytic and numerical methods.
All trivial-phase stable solutions are deformations of the ground state of the 
linear Schr\"odinger equation.  Our results show that a large number of
condensed atoms is sufficient to form a stable, periodic condensate. 
Physically, this implies stability of states near the Thomas--Fermi limit.
\end{abstract}


\begin{multicols}{2}

\section{Introduction}

Recent experiments on dilute--gas Bose-Einstein condensates (BECs) have
generated great interest in macroscopic quantum
phenomena~\cite{ketterle1,dalfovo1} in both the theoretical and experimental
physics community.  Such BECs are experimentally realized when certain gases
are super-cooled below a critical temperature and trapped in electromagnetic
fields~\cite{huang}.  
Many BEC experiments use harmonic confinement.
Recently, however, there has been much interest in sinusoidal 
confinement of repulsive
BECs using standing light waves.  
Such BECs have been used to study phase
coherence~\cite{anderson3,hagley,crawford} 
and matter-wave diffraction~\cite{ovchinnikov1}.
They have also been predicted to apply to quantum logic~\cite{jaksch1,brennen},
matter-wave transport~\cite{choi1}, and matter-wave gratings.
In this paper, we consider the dynamics and stability of repulsive BECs trapped 
in standing light waves. 

A mean--field description for the macroscopic BEC wavefunction is constructed
using the Hartree--Fock approximation~\cite{hartree} and results in the
Gross-Pitaevskii equation~\cite{pitaevskii1,gross1}.  The dimensions of the
BEC play an important role: 1D, 2D, and 3D BECs all behave in a radically
different manner~\cite{petrov1,petrov2}.  In the quasi-1D regime, the
Gross-Pitaevskii equation reduces to the 1D nonlinear Schr\"odinger equation
(NLS) with an external potential.  This regime holds when the transverse
dimensions of the condensate are on the order of its healing length and the
longitudinal dimension is much longer than its transverse
dimensions~\cite{carr15,carr22}.  In this regime the BEC remains phase
coherent and the governing equations are one-dimensional.
This is in contrast to a truly 1D mean-field theory which requires 
transverse dimensions on the order of or less than 
the atomic interaction length.


The recent trapping of a BEC in a hollow blue-detuned laser beam~\cite{bongs1} 
demonstrates that a quasi-1D BEC is experimentally realizable.  A variety of other 
experiments~\cite{ketterle1,key1,dekker1,bongs1,andrews1,close1,matthews1} 
are also modeled by the 1D NLS with an external potential.
Upon rescaling, the governing evolution is given by
\begin{equation}
\label{eqn:NLS}
 i\psi_t = -\frac{1}{2}\psi_{xx} + |\psi|^2 \psi 
        + V(x) \psi \, ,
\end{equation}
where $\psi(x,t)$ is the macroscopic wave function of the condensate and $V(x)$
is an experimentally generated macroscopic potential.   Confinement in a
standing light wave results in $V(x)$ being periodic. In a recent
experiment~\cite{anderson3,hagley}, a shallow harmonic potential was applied in
addition to a standing light wave.   The standing light wave in this case was
sufficiently intense so that the condensate was strongly localized in each
well. This is referred to as the tight-binding regime.  Additionally, the
apparatus was tilted vertically so that gravity caused  tunneling between
wells.  Our theoretical findings consider the complimentary experiment in which
the condensate is free to move between wells. 
With the advent of quasi-1D, 
cylindrical geometries~\cite{bongs1}, additional harmonic confinement is no
longer necessary and the BEC dynamics considered here are applicable.

\begin{figure}[htb]
\centerline{\psfig{figure=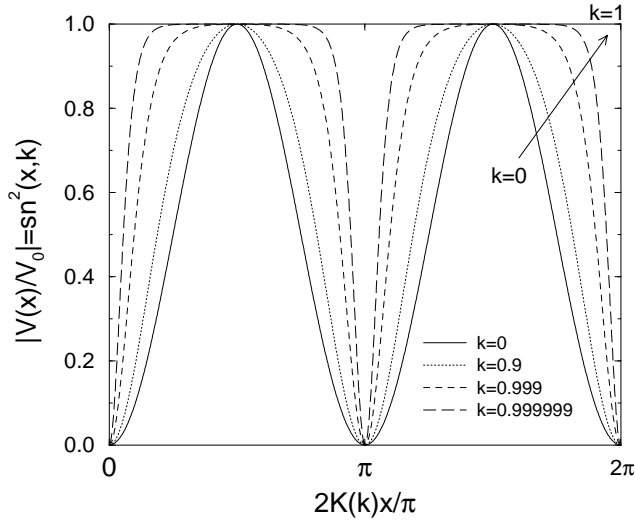,width=83mm,silent=}}
\begin{center}
\begin{minipage}{83mm}
\caption{ \label{fig:Sn_potential} The ${\rm sn}^2(x,k)$ structure of
the potential for varying
values of $k$. Note that the $x$-coordinate has been scaled by the period of the
elliptic function. This period approaches infinity as $k\rightarrow 1$.
Since ${\rm sn}(x,k)$ is periodic in $x$
with period $4K(k)=4\int_0^{\pi/2}{d\alpha}/{\sqrt{1-k^2\sin^2\alpha}}$,
$V(x)$ is periodic in $x$ with period $2K(k)$.} 
\end{minipage}
\end{center}
\end{figure}

To model the quasi-1D confinement produced by a standing light wave,
we use the periodic potential
\begin{equation}
    V(x) = -V_0~ {\rm sn}^2(x,k)
\label{eqn:potential}
\end{equation}
where ${\rm sn}(x,k)$ denotes the Jacobian elliptic sine
function~\cite{abro} with elliptic modulus $0\leq k\leq 1$.  In the
limit $k=0$ the potential is sinusoidal and thus $V(x)$ is a
standing light wave.  For intermediate values (e.g. $k<0.9$) the potential
closely resembles the sinusoidal behavior and thus provides a good
approximation to a standing light wave.  Finally, for $k\rightarrow 1^-$,
$V(x)$ becomes an array of well-separated hyperbolic secant potential barriers
or wells.  The potential is plotted in Fig.~\ref{fig:Sn_potential} for values
of $k=0, 0.9, 0.999$ and $0.999999$.  Only for $k$ very near one (e.g.
$k>0.999$) does the solution appear visibly elliptic.
%
%
%
%
\noindent
The freedom in
choosing $k$ allows us to consider much more general potentials
than considered previously~\cite{barra1,barra2,steel1,steel2} 
and allows for great flexibility 
in considering a wide variety of physically realizable potentials.  

The paper is outlined as follows: in the next section we derive and
consider various properties and limits of
two types of explicit solutions of Eq.~(\ref{eqn:NLS}) with 
(\ref{eqn:potential}).  Section~III develops the analytic framework
for the linear stability properties of the new solutions of Sec.~II.
The stability results are confirmed by numerical computations. In certain
cases, the stability analysis is intractable and we rely solely on
simulations to determine stability.  
%
%
We conclude the paper in Sec.~IV with a brief summary and highlights
of the primary results of the paper and their consequences for
BEC dynamics and confinement.

\section{Stationary Solutions}

Equation~(\ref{eqn:NLS}) with $V(x)=0$ is an integrable equation and many
explicit solutions corresponding to various boundary conditions are known. A
comprehensive overview of these solutions is found in~\cite{belokolos}. 
If $V(x)\neq 0$, the NLS is not
integrable. In this case, only small classes of explicit solutions can most
likely be obtained.  Our choice of potential ~(\ref{eqn:potential}) is
motivated by the form of the stationary solution of the NLS with $V(x)=0$. An
overview of these stationary solutions and their properties is found
in~\cite{carr15}.  At present, we restrict our attention to stationary solutions of
Eq.~(\ref{eqn:NLS}), $i.e.,$ solutions whose time-dependence is restricted to  
\begin{equation} 
\psi(x,t) = r(x)~\exp(-i \omega t+i \theta(x)) \, .  
\label{eqn:ansatz} 
\end{equation} 
If $\theta_x\equiv 0$, then the solution is referred to as having trivial
phase and we choose $\theta(x)=0$.  Substituting the ansatz
Eq.~(\ref{eqn:ansatz}) in Eq.~(\ref{eqn:NLS}) and dividing out the exponential factor
results in two equations: one from the real part and one from the imaginary
part. The second equation can be integrated:  
\begin{equation} 
\theta(x) = c\int_0^x \frac{dx'}{r^2(x')} \, ,
\label{eqn:genphase} 
\end{equation} 
where $c$ is a constant of integration. Note that $\theta(x)$ is a
monotonous function of $x$. Substitution of this result in the remaining
equation gives 
\begin{equation} 
\omega r^4(x)=\frac{c^2}{2}-\frac{r^3(x)r''(x)}{2}+r^6(x)-V_0~{\rm sn}^2(x,k)
r^4(x). 
\label{eqn:ode} 
\end{equation} 
The following subsections describe two classes of solutions of this equation. 

\subsection*{Type A} 

\subsubsection{Derivation}

For these solutions, $r^2(x)$ is a quadratic function of
sn$(x,k)$: 
\begin{equation} 
r^2(x) = A~{\rm sn}^2(x,k)+B.
\label{eqn:quadratic} 
\end{equation} 
Substituting this ansatz in Eq.~(\ref{eqn:ode}) and equating the coefficients of
equal powers of ${\rm sn}(x,k)$ results in relations among the solution
parameters $\omega, c, A$ and $B$ and the equation parameters $V_0$ and $k$.
These are 
\begin{subeqnarray}\label{eqn:parametersA}
\omega &=& \frac{1}{2}\left(1+k^2+3B-\frac{BV_0}{V_0+k^2}\right),\\
c^2 &=& B~\left(1+\frac{B}{V_0+k^2}\right)\left(V_0+k^2+B k^2\right),\\
A &=& V_0+k^2.
\end{subeqnarray}
For a given potential $V(x)$, this solution class has one free parameter
$B$ which plays the role of a constant background level or offset. The freedom
in choosing the potential gives a total of three free parameters: $V_0$, $k$
and $B$. 

The requirements that both $r^2(x)$ and $c^2$ are positive imposes conditions
on the domain of these parameters:
\begin{subeqnarray}\label{eqn:validityA} 
&V_0\geq -k^2,~~ B\geq0,~~~~\mbox{or}&\\
&V_0\leq -k^2,~~ -(V_0+k^2)\leq B \leq
-\left(1+\displaystyle{\frac{V_0}{k^2}}\right) \, .& 
\end{subeqnarray} 
%
%
\begin{figure}[htb]
\centerline{\psfig{figure=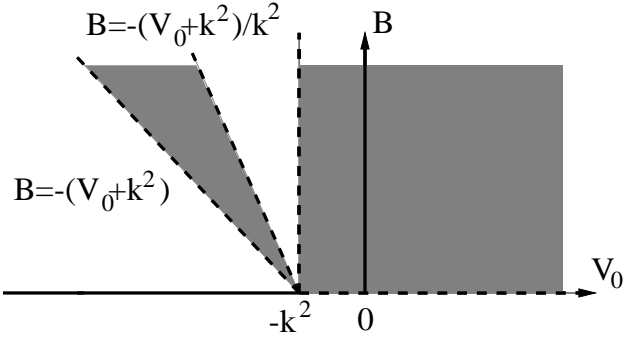,width=83mm,silent=}}
\begin{center}
\begin{minipage}{83mm}
\caption{ \label{fig:validity1} The region of validity of the solutions of Type
A is displayed shaded for a fixed value of $k$. The edges of these regions
correspond to various trivial phase solutions.}
\end{minipage}
\end{center}
\end{figure}
%
%
\noindent The region of validity of these solutions is displayed in
Fig.~\ref{fig:validity1}.

For typical values of $V_0, k$ and $B$, the above equations give rise to
solutions of Eq.~(\ref{eqn:NLS}) which are not periodic in $x$: $r(x)$ is
periodic with period $2K(k)$, whereas $\exp(i \theta(x))$ is periodic with
period $T=\theta^{-1}(2\pi)$. In general these two periods $2K(k)$ and $T$ are
not commensurable.  Thus, requiring periodic solutions results in another condition,
namely $2K(k)/T=p/q$, for two positive integers $p$ and $q$. The most
convenient way to express this phase quantization condition is to assume the
potential ($i.e.,$ $V_0$ and $k$) is given, and to consider values of $B$ for which
the quantization condition is satisfied. Introducing $\beta=B/(V_0+k^2)$, we
find
\begin{equation}\label{eqn:genquant}
\pm\frac{\sqrt{\beta(1+\beta)(1+k^2\beta)}}{\pi}\int_0^{K(k)}\frac{dx}{{\rm
sn}(x,k)^2+ \beta}=\frac{p}{q}. 
\end{equation}
This equation is solved for $\beta$, after which $B=\beta (V_0+k^2)$. For
numerical simulations, the number of periods of the potential is set. This
determines $q$, limiting the number of solutions of Eq.~(\ref{eqn:genquant}). 
Solutions with the same periodicity as the potential require $p/q=1$. 

Note that solutions of Type A reduce to stationary solutions of Eqs.~(\ref{eqn:NLS})
and (\ref{eqn:potential}) with $V_0=0$. Furthermore, all stationary solutions
of the integrable equation are obtained as limits of solutions of Type A.

\subsubsection{Limits and Properties}

The properties of these solutions are best understood by considering their
various limit cases. 

\begin{figure}[htb]
\centerline{\psfig{figure=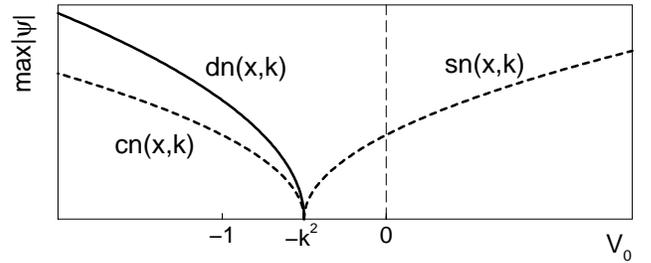,width=83mm,silent=}}
\begin{center}
\begin{minipage}{83mm}
\caption{ \label{fig:bif} The amplitude of the trivial phase solutions of Type A
versus the potential strength $V_0.$}
\end{minipage}
\end{center}
\end{figure}

{\bf The trivial phase case:} The solutions of Type A have trivial phase when
$c=0$. Since $c^2$ has three factors which are linear in $B$ (see
Eq.~(\ref{eqn:parametersA})), there are three choices of $B$ for which this occurs:
$B=0$, $B=-(V_0+k^2)$ and $B=-(V_0+k^2)/k^2$. These possibilities are three of
the four boundary lines of the region of validity in Fig.~\ref{fig:validity1}.
Note that the remaining boundary line ($V_0=-k^2$) corresponds to $r^2(x)=B$,
which gives rise to a plane wave solution. Using Jacobian elliptic
function identities~\cite{abro}, one finds that the three other boundary lines give rise
to simplified solution forms: $B=0$ gives 
\begin{equation}\label{eqn:sn}
r(x)=\sqrt{V_0+k^2}~{\rm sn}(x,k), \,\, ~\omega=\frac{1+k^2}{2}.
\end{equation}
$B=-(V_0+k^2)$ gives
\begin{equation}\label{eqn:cn}
r(x)=\sqrt{-(V_0+k^2)}~{\rm cn}(x,k), \,\, ~\omega=\frac{1}{2}-V_0-k^2,
\end{equation}
where ${\rm cn}(x,k)$ denotes the Jacobian elliptic cosine function. 
Lastly, $B=-(V_0+k^2)/k^2$ gives
\begin{equation}\label{eqn:dn}
r(x)=\frac{\sqrt{-(V_0+k^2)}}{k}~{\rm dn}(x,k), \,\,
~\omega=-1-\frac{V_0}{k^2}+\frac{k^2}{2},
\end{equation}
where ${\rm dn}(x,k)$ denotes the third Jacobian elliptic function.
Solution~(\ref{eqn:sn}) is valid for $V_0\geq-k^2$, whereas the other two 
solutions~(\ref{eqn:cn}) and (\ref{eqn:dn}) are valid for $V_0\leq-k^2$. 
The amplitude of
these solutions as a function of potential strength $V_0$ is shown in
Fig.~\ref{fig:bif}. 

%
%

%
\begin{figure}[htb]
\centerline{\psfig{figure=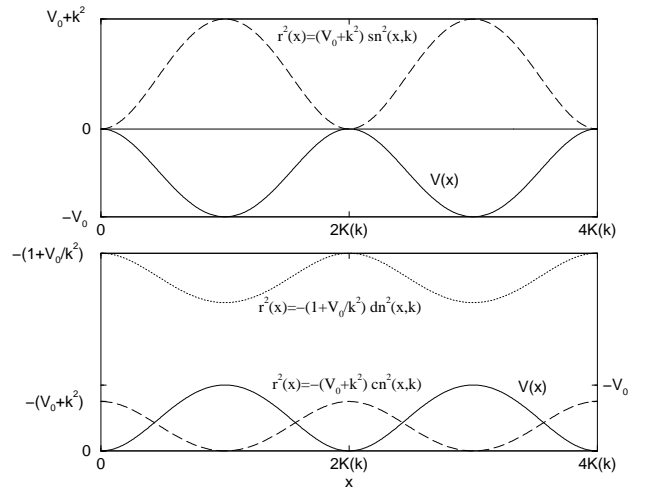,width=83mm,silent=}}
\begin{center}
\begin{minipage}{83mm}
\caption{\label{fig:triv1} Trivial phase solutions for $k=0.5$. $V(x)$ is
indicated with a solid line. For the top figure $V_0=1$. For the bottom figure
$V_0=-1$.}
\end{minipage}
\end{center}
\end{figure}
%

%
\begin{figure}[htb]
\vspace*{6mm}
\centerline{\psfig{figure=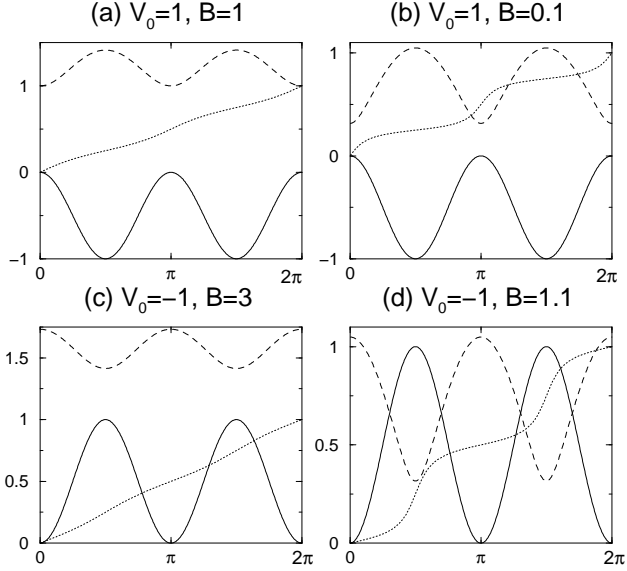,width=83mm,silent=}}
\begin{center}
\begin{minipage}{83mm}
\caption{\label{fig:trig} Phase and amplitude of the 
trigonometric solutions. For all these figures,
the solid line denotes $V(x)$, the dashed line is $r(x)$ and the dotted line is
$\theta(x)/(2\pi)$. Note that $\theta(x)$ becomes piecewise constant, as $B$
approaches the boundary of the region of validity. Far away from this boundary,
$\theta(x)$ is essentially linear.}
\end{minipage}
\end{center}
\end{figure}
%

Both ${\rm cn}(x,k)$ and ${\rm sn}(x,k)$ have zero average as functions of $x$
and lie in $[-1,1]$. On the other hand, ${\rm dn}(x,k)$ has nonzero average.
Its range is $[\sqrt{1-k^2},1]$. Furthermore, ${\rm cn}(x,k)$ and ${\rm
sn}(x,k)$ are periodic in $x$ with period $4K(k)$, whereas ${\rm dn}(x,k)$ is
periodic with period $2K(k)$. These properties matter greatly for the stability
analysis, as will be seen in Section 3. 
Some solutions with trivial phase are shown in Fig.~\ref{fig:triv1}.

{\bf The trigonometric limit:} In the limit $k\rightarrow 0$, the elliptic
functions reduce to trigonometric functions and
$V(x)=-V_0~\sin^2(x)=(V_0/2)~\cos(2x)-V_0/2$. Then 
\begin{equation}\label{eqn:trig}
r^2(x)=V_0~\sin^2(x)+B,~~~\omega=\frac{1}{2}+B.
\end{equation}
In this case, the phase integral Eq.~(\ref{eqn:genphase}) results in 
\begin{equation}\label{trig:phase}
\tan(\theta(x))=\pm\sqrt{1+V_0/B}~\tan(x).
\end{equation}
Note that this formula guarantees that the resulting solution is periodic with
the same period as the potential, so no phase quantization is required. In the
trigonometric limit, the wedge between the two regions of validity in
Fig.~\ref{fig:validity1} disappears. This is no surprise, as in this limit,
${\rm dn}(x,k)\rightarrow 1$, and the third trivial phase solution reduces
to a plane wave solution. The cornerpoint of the region of validity also moves
to the origin. Some trigonometric solutions are illustrated in Fig.~\ref{fig:trig}.

\begin{figure}[htb]
\vspace*{6mm}
\centerline{\psfig{figure=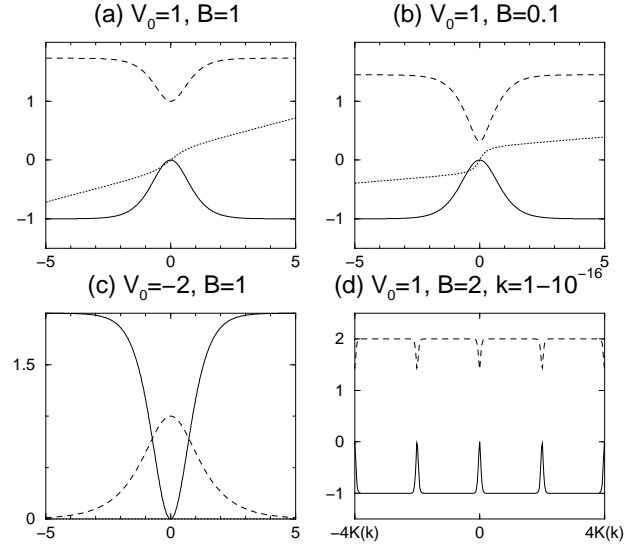,width=83mm,silent=}}
\begin{center}
\begin{minipage}{83mm}
\caption{\label{fig:solitons} Solutions with k=1 (a,b,c) or $k\rightarrow 1$
(d). The solid line denotes $V(x)$, the dashed line is
$r(x)$ and the dotted line is $\theta(x)/2\pi$. In (d), a value of $k=1-10^{-16}$
was used.}
\end{minipage}
\end{center}
\end{figure}

{\bf The solitary wave limit:} $k=1$.  In this limit the elliptic functions
reduce to hyperbolic functions. Specifically, ${\rm sn}(x,k)=
\tanh(x)$. Hence in this limit, the potential has only a
single well or a single peak. Then $V_0<0$ gives rise to a repulsive potential,
whereas $V_0>0$ gives rise to an attractive potential: $V(x)=-V_0~\tanh^2(x)$.
In this case the phase $\theta(x)$ of Eq.~(\ref{eqn:genphase}) can be calculated
explicitly: 
\begin{subeqnarray}\label{eqn:gensoliton} 
r^2(x)&=&(V_0+1)~\tanh^2(x)+B,\\
\theta(x)\!\!&=&\!\!\sqrt{\frac{B}{V_0+1}}x\!+\!\arctan\!\!\left(\!
\sqrt{\frac{V_0+1}{B}}\tanh(x)\!\right)\!,~~~~~
\end{subeqnarray} 
which is valid for $V_0>-1$ and $B>0$. This solution is a
stationary solitary wave of depression on a positive background. It
is a deformation of the gray soliton solution of the NLS
equation with repulsive nonlinearity. Note that these solutions can exist with
an attractive potential provided $-1<V_0<0$. Two solutions with repulsive
potential are illustrated in Fig.~\ref{fig:solitons}a-b. 
Another solution is valid for $B=-(V_0+1)>0$: 
$r(x)=\sqrt{-(V_0+1)}~{\rm sech}(x)$ and 
$\theta(x)=0$. 
This solution represents a stationary elevated solitary wave. It is
reminiscent of the bright soliton solution of the NLS equation with attractive
nonlinearity. This solution is shown in Fig.~\ref{fig:solitons}c.  A surprising
consequence of considering Eq.~(\ref{eqn:NLS}) is that the potential strength
$V_0$ acts as a switch between the equation with repulsive and attractive
nonlinearity, as illustrated by these solitary wave solutions.  

Understanding the solitary wave limit facilitates the understanding of what
occurs for $k\rightarrow 1$. In this case the solutions of Type A
reduce to a periodic train of solitons with exponentially small interactions
as illustrated in Fig.~\ref{fig:solitons}d.

%

\subsection*{Type B}

\subsubsection*{1.~Derivation}

For these solutions, $r^2(x)$ is linear in ${\rm cn}(x,k)$ or ${\rm dn}(x,k)$.
First we discuss the solution with ${\rm cn}(x,k)$. The quantities associated
with this solution will be denoted with a subindex 1. The quantities associated
with the ${\rm dn}(x,k)$ solution receive a subindex 2. 

Substituting
\begin{equation}\label{eqn:linear1}
r_1^2(x)=a_1~{\rm cn}(x,k)+b_1 \, , 
\end{equation}
in Eq.~(\ref{eqn:ode}) and equating different powers of ${\rm cn}(x,k)$ 
gives the relations:
\begin{subeqnarray}\label{eqn:parametersB1}
V_0&=&-\frac{3}{8}k^2 \, ,\\
\omega_1&=&\frac{1}{8}(1+k^2)+\frac{6a_1^2}{k^2} \, ,\\
c^2_1&=&\frac{a_1^2}{4k^6}(16a_1^2-k^4)(16a_1^2+k^2-k^4) \, ,\\
b_1&=&\frac{4a_1^2}{k^2} \, .
\end{subeqnarray}
The class of potentials Eq.~(\ref{eqn:potential}) is restricted by the first of
these relations so that $V_0$ is in the narrow range
$-3k^2/8\leq V_0 \leq0$. The solution class now depends on one free amplitude
parameter $a_1$ and the free equation parameter $k$. 

The region of validity of this solution is, as before, determined by the
requirements $c_1^2\geq 0$ and $r_1^2(x)\geq 0$:
\begin{equation}\label{eqn:validityB1}
|a_1|\geq\frac{k^2}{4}.
\end{equation}
The period of $r_1(x)$ is twice the period of the potential. Requiring
periodicity in $x$ of this first solution of Type B gives 
\begin{equation}\label{eqn:genquant1}
\pm\frac{\sqrt{(\beta_1^2 \!- \!k^2)(\beta_1^2\!+\!1\!-\!k^2)}}{4\pi}
\!\! \int_0^{2K(k)} \!\!\!\!\!\!\!
\frac{dx}{4\beta_1\!+\!k~{\rm cn}(x,k)} \!=\! \frac{p}{q}.
\end{equation}
For given $k$ and integers $p$, $q$, this equation is
solved for $\beta_1$, from which $a_1=\beta_1 k/4$. 

The ${\rm dn}(x,k)$ solutions are found by substituting
\begin{equation}\label{eqn:linear2}
r_2^2(x)=a_2~{\rm dn}(x,k)+b_2,\\
\end{equation}
in Eq.~(\ref{eqn:ode}). Equating different powers of ${\rm dn}(x,k)$
imposes the following constraints on the parameters:
\begin{subeqnarray}\label{eqn:parametersB2}
V_0&=&-\frac{3}{8}k^2 \, ,\\
\omega_2&=&\frac{1}{8}(1+k^2)+6a_2^2 \, ,\\
c^2_2&=&\frac{a_2^2}{4}(16a_2^2-1)(16a_2^2+k^2-1) \, ,~\\
b_2&=&4a_2^2 \, .
\end{subeqnarray}
The class of potentials (\ref{eqn:potential}) is restricted as for the previous
solution by the first of
these relations. The solution class again depends on one free amplitude
parameter $a_2$ and the free equation parameter $k$. 

The region of validity of this solution is once more determined by the
requirements $c_2^2\geq 0$ and $r_2^2(x)\geq 0$: 
\begin{equation}\label{eqn:validityB2}
|a_2|\geq\frac{1}{4}~~\mbox{or}~~0\leq a_2\leq\frac{\sqrt{1-k^2}}{4}. 
\end{equation}

The period of $r_2(x)$ is equal to the period of the potential. Requiring
periodicity in $x$ of this second solution of Type B gives  
\begin{equation}\label{eqn:genquant2} 
\pm\frac{\sqrt{(16
a_2^2\!-\!1)(16a_2^2\!+\!k^2\!-\!1)}}{\pi} \!\! \int_0^{K(k)} \!\!\!\!\!\!\! 
 \frac{dx}{4a_2 \!+\!{\rm
dn}(x,k)}\!=\! \frac{p}{q}.
\end{equation} 
For given $k$ and integers $p$, $q$, this equation needs to
be solved to determine $a_2$. 

In contrast to solutions of Type A, solutions of Type B do not have a
nontrivial trigonometric limit. In fact, for solutions of Type B,
this limit is identical to the limit in which the potential strength
$V_0=-3k^2/8$ approaches zero. Thus it is clear that the solutions of Type B
have no analogue in the integrable nonlinear Schr\"{o}dinger equation. However,
other interesting limits do exist. 

\subsubsection*{2.~Limits and Properties}


{\bf The trivial phase case:} Trivial phase corresponds to $c=0$. This occurs
precisely at the boundaries of the regions of validity. For the first solution
of Type B, there are two possibilities: $a_1=k^2/4$ or $a_1=-k^2/4$. By replacing
$x$ by $x+2K(k)$, one sees that these two possibilities are completely
equivalent, so only the first one needs to be considered:
\begin{equation}\label{eqn:trivphaseB1}
r_1^2(x)=\frac{k^2}{4}(1+{\rm cn}(x,k))~,~~~\omega_1=\frac{1}{8}+\frac{k^2}{2}. 
\end{equation}
For the second solution, there are four possibilities: $a_2=1/4$, $a_2=-1/4$,
$a_2=0$ and $a_2=\sqrt{1-k^2}/4$. The third one of these results in a zero
solution. The others give interesting trivial phase solutions. For $a_2=1/4$, 
\begin{equation}\label{eqn:trivphaseB2}
r_2^2(x)=\frac{1}{4}(1+{\rm dn}(x,k))~,~~~\omega_2=\frac{1}{2}+\frac{k^2}{8}. 
\end{equation}
The case $a_2=-1/4$ gives 
\begin{equation}\label{eqn:trivphaseB3}
r_2^2(x)=\frac{1}{4}(1-{\rm dn}(x,k))~,~~~\omega_2=\frac{1}{2}+\frac{k^2}{8}. 
\end{equation}
Finally, for $a_2=\sqrt{1-k^2}/4$,
\begin{equation}\label{eqn:trivphaseB4}
r_2^2(x)\!=\!\frac{\sqrt{1\!-\!k^2}}{4}({\rm dn}(x,k)\!+\!\sqrt{1\!-\!k^2}),
~~\omega_2\!=\!\frac{1\!-\!k^2}{2}. 
\end{equation}
%
%
\begin{figure}[htb]
\centerline{\psfig{figure=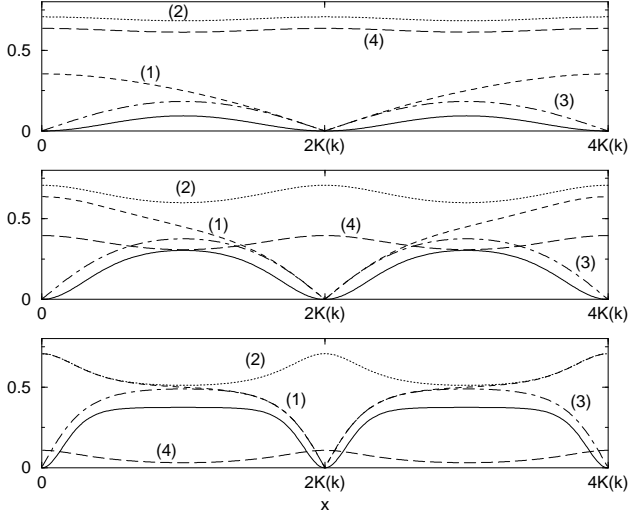,width=83mm,silent=}}
\begin{center}
\begin{minipage}{83mm}
\caption{\label{fig:trivphase2} Solutions of Type B with trivial phase. The 
figures correspond to, from top to bottom, $k=0.5$, $k=0.9$ and $k=0.999$. The
potential is indicated with a solid line. The other curves are: (1) $|r_1(x)|$
with $a_2=k^2/4$, (2) $r_2(x)$ with $a_2=1/4$, (3) $|r_2(x)|$ with $a_2=-1/4$
and (4) $r_2(x)$ with $a_2=\sqrt{1-k^2}/4$.}
\end{minipage}
\end{center}
\end{figure}
%
%
\begin{figure}[htb]
\centerline{\psfig{figure=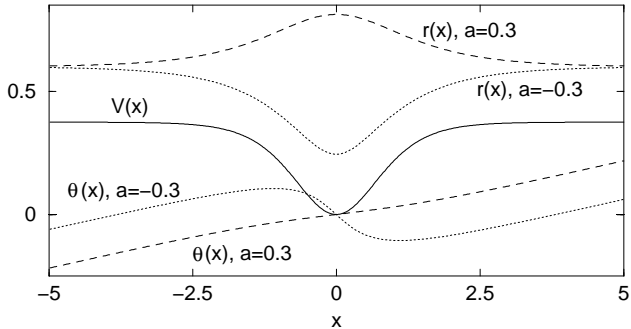,width=83mm,silent=}}
\begin{center}
\begin{minipage}{83mm}
\caption{\label{fig:solitonsB} 
Solitary wave solutions of Type B. The potential is indicated with a solid
line. The dashed line solution corresponds to $a=0.3$, the dotted line to
$a=-0.3$.}
\end{minipage}
\end{center}
\end{figure}
%
%
\noindent
These solutions are shown in Fig.~\ref{fig:trivphase2}.

{\bf The solitary wave limit:} In this limit $V_0=-3/8$ and the potential is
$V(x)=3 \tanh(x)^2/8$. Since both ${\rm cn}(x,k) = {\rm sech}(x)$ 
and ${\rm dn}(x,k)= {\rm sech}(x)$ when $k=1$,
$r_1(x)=r_2(x)$ in the solitary wave limit. Their ranges of validity also share
the same limit: $|a|\geq 1/4$ with $a=a_1=a_2$. The phase can be calculated
explicitly and the solitary wave solution of Type B is 
\begin{subeqnarray}\label{eqn:solitonB} 
r^2(x)&\!\!\!=\!\!\!&4a^2+a~{\rm sech}(x),\\
\theta(x)&\!\!=\!\!\!&\pm \! \frac{x\sqrt{16a^2\!-\!1}}{2}\!
   \mp \!\arctan\left(\!\sqrt{\frac{4a\!-\!1}{4a\!+\!1}}
   \tanh\frac{x}{2}\!\right) . \,\,\,\,\,\,\,\,\,\,\,\,\,
\end{subeqnarray} 
The region of validity consists of two separated regions: $a\geq 1/4$ and
$a\leq -1/4$. In the first region, the solution is a stationary elevated
solitary wave with a constant background $4a^2$. In the second region $a\leq
-1/4$ and the solution is a stationary solitary wave of depression with a
constant background $4a^2$. These solitary wave solutions are illustrated in
Fig.~\ref{fig:solitonsB}

As for the solutions of type A, the solitary wave limit gives an idea of the
behavior of the solution for values of $k\rightarrow 1$, where the solution
behaves as a periodic array of solitons with exponentially small interactions.

\section{Stability}\label{sec:stability}

We have found a large number of new solutions to the
governing Eqs.~(\ref{eqn:NLS}) and (\ref{eqn:potential}).  However, 
only solutions that are stable can be observed in experiments.  In
this section, we consider the stability of the different solutions.
Both analytical and numerical results are presented for the solutions
with trivial phase.  In contrast, 
only numerical results are presented for the nontrivial phase
cases.

We first consider the linear stability of the
solution~(\ref{eqn:ansatz}).  To do so, consider perturbations 
of the exact solutions of the form
\begin{equation}\label{eqn:perturb} 
  \psi(x,t)=(r(x)+\epsilon\phi(x,t)) \exp [{i(\theta(x)-\omega t)}]
\end{equation}
where $\epsilon \ll 1$ is a small parameter.
Collecting terms at $O(\epsilon)$ gives the linearized
equation. In terms of the real and imaginary parts   
${\bf U}=(U_1,U_2)^t=(Re[\phi],Im[\phi])^t$ the linearized evolution 
is given by: 
\begin{equation} 
\label{eqn:linearized}
   {\bf U}_t=JL{\bf U}=J\pmatrix{L_+ & -S \cr S & L_-} {\bf U},
\end{equation}
where 
\begin{subeqnarray}
L_+ &= & -\demi \left(\partial_x^2-\frac{c^2}{r^4(x)} \right)+3r^2(x)
+V(x)-\omega,~~~~\\
L_- & = & -\demi \left(\partial_x^2-\frac{c^2}{r^4(x)} \right)+r^2(x)
+V(x)-\omega,\\
S &=& \frac{c}{r(x)}\partial_x\frac{1}{r(x)},
\end{subeqnarray}
and $J=\pmatrix{0&-1\cr 1&0}$ is a skew--symmetric matrix. The operators
$L$, $L_+$ and $L_-$ are Hermitian while $S$ is anti--Hermitian.
Considering solutions of the form ${\bf U} (x,t) = {\hat{\bf U}}(x)\exp
(\lambda t)$
gives the eigenvalue problem
\begin{equation}
\label{eqn:keith}
  \cL \hat{\bf U}=\lambda \hat{\bf U},
\end{equation}
where $\cL=JL$ and $\lambda$ is complex.  If all $\lambda$ are imaginary,
then linear stability is established.  In contrast, if there is at
least one eigenvalue with a positive real part, then instability results.
Using the phase invariance $\psi\mapsto e^{i\gamma}\psi$ of 
Eq.~(\ref{eqn:NLS}), Noether's theorem~\cite{classical} gives 
\begin{equation}
\label{eqn:null}
  \cL \pmatrix{0\cr r(x)}=0,
\end{equation}
which implies that $L_- r(x)=0.$  Thus $\lambda=0$ is in
the spectrum of $L_-$.  
For general solutions of the form (\ref{eqn:ansatz}),
determining the spectrum of the associated linearized 
eigenvalue problem (\ref{eqn:linearized}) is beyond the scope of current 
methods.  
However, some
cases of trivial phase solutions $(c=0)$ are amenable to analysis. 

The Hermitian operators $L_\pm$ are periodic Schr\"{o}dinger 
operators and thus the spectra of these operators is real and 
consists of bands of continuous spectrum contained in 
$[\lambda_\pm,\infty)$\cite{classical}. Here $\lambda_\pm$ denote 
the ground state eigenvalues of $L_\pm$ respectively. They are given by
\begin{equation}
  \lambda_\pm=\inf\limits_{\|\phi\|=1} \left< \phi|L_\pm |\phi 
     \right> \label{Linf} \, ,
\end{equation}
where $\|\phi\|^2=\left<\phi|\phi\right>$.   
From the relation $L_+=L_-+2r^2(x)$ it follows that $\lambda_+>\lambda_-$.
Also $\lambda_-\leq 0$ since $\lambda=0$ is an eigenvalue of $L_-$.

If $\lambda_+ > 0$, then $L_+$ is positive, so we can
define the positive square root,  $L_+^\demi$, via the spectral
theorem\cite{classical}, and hence the Hermitian operator 
$H=L_+^\demi L_- L_+^\demi$ can be constructed. The eigenvalue 
problem for $\cL$ in Eq.~(\ref{eqn:keith}) is then equivalent to 
\begin{equation}\label{eqn:H}
  (H+\lambda^2)\varphi=0,  
\end{equation}
with $\varphi=L_+^\demi\hat{U_1}$. 
Denote the left-most point of the spectrum of $H$
by $\mu_0$.  If $\mu_0\geq 0$ then $\lambda^2<0$ and
the eigenvalues of $\cL$ are imaginary and linear stability
results. Since $H=L_+^\demi L_-
L_+^\demi$ and $L_+^\demi$ is positive, $\mu_0\geq 0$ if and only if $L_-$ is
non-negative. 
In contrast, if $\mu_0<0$ then $\lambda^2>0$ and
$\cL$ has at least one pair of real eigenvalues with opposite
sign.  This shows the existence of a growing mode leading
to instability of the solution.

Three distinct cases are possible for linear stability
\begin{itemize}
\item If $r(x)>0$ then $r(x)$ is the ground state of $L_-$~\cite{classical}, 
and Eq.~(\ref{eqn:null}) implies $\lambda_-=0$ and hence $\lambda_+>0$.
Thus the solution (\ref{eqn:ansatz}) is linearly stable.
\item If $r(x)$ has a zero, it is no longer the ground
state~\cite{classical} and $\lambda_-<0$.  Thus  
there exists a $\psi_0$ such that $\left<\psi_0 | L_- |\psi_0 \right> <0$.
If in addition $\lambda_+>0$, then we can construct 
$\phi=L_+^{-\demi}\psi_0/\|L_+^{-\demi}\psi_0\|$ which 
gives $\left< \phi | H |\phi \right><0$.  Hence
$\mu_0<0$ and $\cL$ has positive, real eigenvalues so that
the solution (\ref{eqn:ansatz}) is linearly
unstable.  
\item For $\lambda_-$ and $\lambda_+$ both negative the
situation is indefinite and our methods are insufficient
to determine linear stability or instability.  
\end{itemize}
In what follows,  
these results are applied to the Type A and B trivial phase solutions
constructed in the preceding section.  ~Specifically,~ we
construct~
the~ operators~~$L_-$ 
%
%
\begin{figure}[t]
\centerline{\psfig{figure=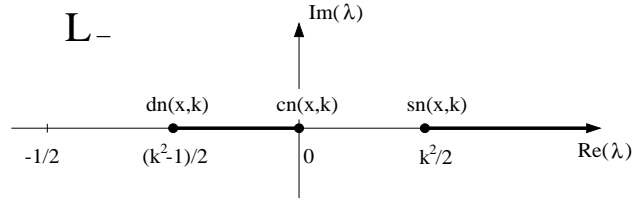,width=83mm,silent=}}
\begin{center}
\begin{minipage}{83mm}
\caption{The spectrum of $L_-$ for the Type A cn$(x,k)$ trivial phase solution.}
\label{fig:speccn}
\end{minipage}
\end{center}
\end{figure}
%
%
\noindent
and $L_+$ for each solution, which allows us to
use one of the above criteria.  The analytical results are accompanied
by direct computations on the nonlinear governing Eqs.~(\ref{eqn:NLS}) and
(\ref{eqn:potential}).   For all computational simulations, twelve 
spatial periods are used.  However, to better illustrate the dynamics,
typically four spatial periods are plotted.  Moreover, all computations
are performed with white noise included in the initial data.

\subsection{Trivial Phase:  Type A}

\subsubsection{cn$(x,k)$}

For the cn$(x,k)$ solution the $L_\pm$ operators are 
\begin{subeqnarray}\label{eqn:speccn}
 L_+&\!=\!&-\demi\partial_x^2-(2V_0+3k^2)\cn^2(x,k)+k^2-\demi \,\,\,\,\, \\
 L_-&\!=\!&-\demi\partial_x^2-k^2\cn^2(x,k)+k^2-\demi \, ,
\end{subeqnarray}
with $V_0<-k^2.$ Note that $L_-$, which is independent of $V_0$, is the
classical 1-gap Lam\'e operator  
~\cite{belokolos}. The
spectrum of $L_-$ can be calculated explicitly. The ground state eigenvalue is
$\lambda_-=(k^2-1)/2$ with associated eigenfunction dn$(x,k)$.  The elliptic 
functions
$\cn(x,k)$ and $\sn(x,k)$ are also eigenfunctions of $L_-$. They are the first
and second excited state and have eigenvalue 0 and $k^2/2$ respectively. 
These are the only eigenvalues and the spectrum consists of the bands 
$[(k^2-1)/{2},0]\cup[{k^2}/{2},\infty)$. The spectrum is illustrated in 
Fig.~\ref{fig:speccn}.

Since $\dn(x,k)>0$ and $\lambda_-=({k^2-1})/{2}<0$
the arguments of the previous section imply that
the cn$(x,k)$ wave is unstable whenever the operator 
$L_+>0$.  It is clear from Eq.~(\ref{eqn:speccn}a) that $L_+$ is positive 
if $V_0<-(k^2 + 1/4)$ and $k^2 > 1/2$. Thus, the cn$(x,k)$ wave is unstable
for parameter values in this region.
Moreover, this region can be enlarged to $V_0<-(k^2+1)/2$ and $k^2>1/2$ by observing that
the ground state eigenvalue of an operator 
${L_0 + \gamma L_1}$ is a convex function of $\gamma$:
$\lambda=\Lambda(\gamma)$. 
This follows from the fact that the ground state eigenvalue 
is the minimizer of the Rayleigh quotient. Let $\alpha\in [0,1]$, then
%
%
\begin{figure}[t]
\centerline{\psfig{figure=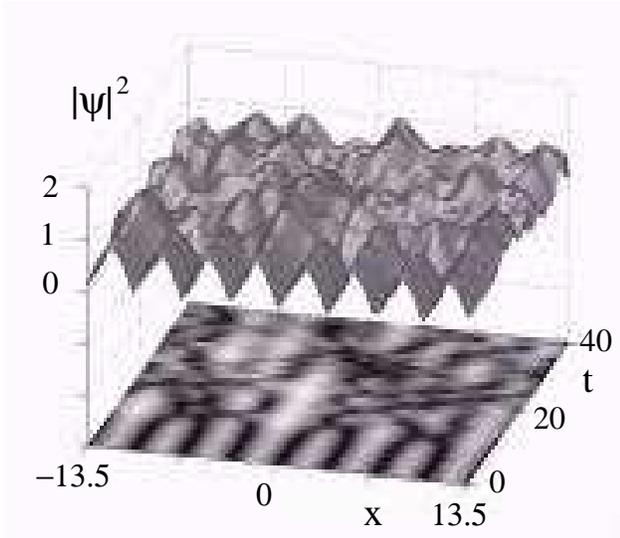,width=83mm,silent=}}
\begin{center}
\begin{minipage}{83mm}
\caption{Unstable evolution of a Type A cn$(x,k)$ solution
given by Eq.~(\ref{eqn:cn}) over 40 time units with $k=0.5$ and
$V_0=-1.0$.}
\label{fig:trivialCN}
\end{minipage}
\end{center}
\end{figure}
%
%
%
\begin{figure}[t]
\centerline{\psfig{figure=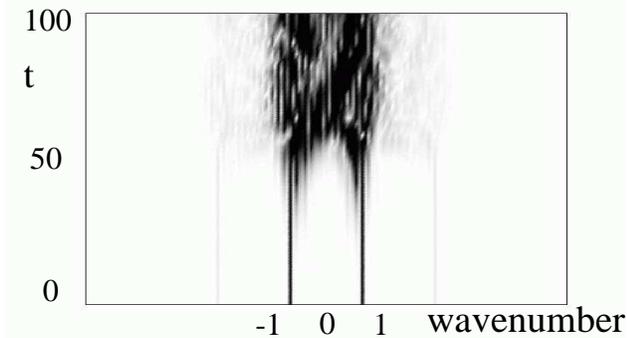,width=83mm,silent=}}
\begin{center}
\begin{minipage}{83mm}
\caption{Wavenumber spectrum evolution of a Type A cn$(x,k)$  
solution given by Eq.~(\ref{eqn:cn}) over 100 time units with 
$k=0.5$ and $V_0=-0.55$.  The modal evolution 
shows the band of unstable modes which result from starting
with the unstable cn$(x,k)$ solution.  This shows that the 
instability
occurs in a neighborhood of the dominant wavenumber of the stationary 
solution.}
\label{fig:spec-trivial}
\end{minipage}
\end{center}
\end{figure}
%
%
\begin{eqnarray}\nonumber
\Lambda\!\!\!\!\!&&\!\!(\alpha\gamma_1+(1-\alpha)\gamma_2)\\\nonumber
&=&\inf\limits_{\|\phi\|=1} \left<\phi|\alpha(L_0+\gamma_1
L_1)+(1-\alpha)(L_0+\gamma_2 L_1)|\phi\right>\\\nonumber
&\geq&\alpha \!\!\inf\limits_{\|\phi\|=1}\!\!\! \left<\phi|L_0\!+\!\gamma_1
L_1|\phi\right>+(1\!-\!\alpha)\!\!\!\inf\limits_{\|\phi\|=1}\!\!\! 
\left<\phi|L_0\!+\!\gamma_2 L_1|\phi\right>\\
&=&\alpha\Lambda(\gamma_1)+(1-\alpha)\Lambda(\gamma_2).
\end{eqnarray} 
Now consider the ground state eigenvalue $\lambda_+=\Lambda_+(V_0)$ 
and note that $\Lambda_+(-k^2) = (k^2-1)/2$ 
and $\Lambda_+(-3k^2/2) = k^2-1/2$. The line through these two 
points is given by $\Lambda_+(V_0)=-V_0 - (1+k^2)/2$, so by convexity 
$\Lambda_+(V_0)\ge -V_0 - (1+k^2)/2$ for $V_0 \in [-3k^2/2,-k^2]$.
Thus, $\lambda_+=\Lambda(V_0) \ge 0$ if $V_0 \le -(1+k^2)/2$. 

If $k^2<1/2$, less is known. However 
the results of Weinstein and Keller~\cite{WK} show that
the ground state eigenvalue grows as $\lambda_+\approx 
(2(1-k^2)|V_0|)^{1/2} + k^2- {1}/{2}$ for $-V_0\gg 1$. 
Hence for $k^2 \leq 1/2$ instability occurs for sufficiently negative $V_0$.   

The most unstable modes of the cn$(x,k)$ solution 
can be determined perturbatively when $\varepsilon=-2(V_0+k^2)\ll 1$. This 
corresponds to a solution with small amplitude. Since 
$L_+ = L_- + \varepsilon \cn^2(x,k)$, it follows that $L_+$ is not necessarily
positive, disallowing the construction of $H$ in Eq.~(\ref{eqn:H}).  
%
%
However, from Eq.~(\ref{eqn:keith}), $L_+L_- \hat{U_2}=-\lambda^2 \hat{U_2}$,
which offers an alternative to Eq.~(\ref{eqn:H}) to calculate the spectrum of
Eq.~(\ref{eqn:keith}). Let $\lambda=i \nu+\varepsilon \lambda_1$ and
$\hat{U_2}=\phi_\nu+\varepsilon \phi_1$, where $\nu$ is an eigenvalue of $L_-$
and $\phi_\nu$ is its associated normalized eigenfunction. Then a first order
calculation using time-independent perturbation theory gives
\begin{equation}
\lambda^2=-\nu^2-\varepsilon \nu \left<\phi_\nu|\cn^2(x,k)|\phi_\nu\right>.
\end{equation}
Thus, $\lambda^2>0$ only if $-\varepsilon 
\left<\phi_\nu|\cn^2(x,k)|\phi_\nu\right> < \nu < 0$. Hence, only modes
$\phi_\nu$ with $\nu$ in this band near zero are unstable. 
%
%
For these unstable modes, the eigenfunction $\phi_\nu$ is approximately the 
zero mode $\cn (x,k)$. Thus the onset of instability in the Fourier domain 
occurs near the wavenumbers of the $\cn(x,k)$ solution. 
%
%
This is characteristic of a modulational instability 

   
To illustrate this instability, we display in Fig.~\ref{fig:trivialCN} the
evolution of a cn$(x,k)$ solution over the time interval $t \in [0,40]$ for
$V_0=-1.0$ and $k=0.5$.  The solution goes quickly unstable with the
instability generated near the first wavenumber.  This agrees with the
analytical prediction.  It is illustrated in the evolution of the wavenumber
spectrum in Fig.~\ref{fig:spec-trivial}.  Here a close-up of the spectrum near
wavenumber one is shown.  This shows that the instability indeed
occurs in a neighborhood of the dominant wavenumber of the stationary solution.

\subsubsection{sn$(x,k)$}

For the sn$(x,k)$ solutions the $L_\pm$ operators are given by
\begin{subeqnarray}
L_+ &\!=\!& -\frac{1}{2}\partial_{xx} + (3 k^2 + 2 V_0)
\sn^2(x,k)-\frac{1+k^2}{2},~~~\\
L_- &\!=\!& -\frac{1}{2}\partial_{xx} + k^2 \sn^2(x,k) - \frac{1+k^2}{2}.
\end{subeqnarray}
Again $L_-$ is a 1-gap Lam\'e operator, differing from $L_-$
for the cn$(x,k)$ solution only by a constant. The spectrum is
given by $[-{k^2}/{2},-{1}/{2}]\cup[0,\infty)$.  It again follows from
the work of Weinstein and Keller~\cite{WK} that for sufficiently large values 
of $V_0$
the ground state eigenvalue of $L_-$ is approximately given by
\begin{equation}
\lambda_+ \approx (2V_0)^\frac{1}{2} - \frac{1+k^2}{2}
\end{equation}
%
%
\begin{figure}[t]
\centerline{\psfig{figure=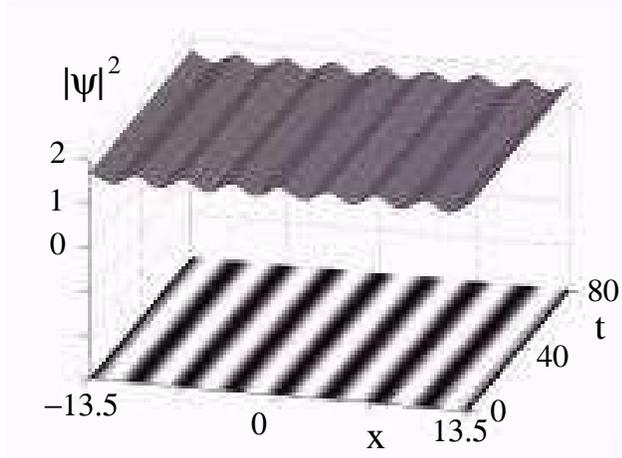,width=83mm,silent=}}
\begin{center}
\begin{minipage}{83mm}
\caption{Stable evolution of the Type A dn$(x,k)$ solutions given
by Eq.~(\ref{eqn:dn}) over 80 time units with $k=0.5$ and $V_0=-1.0$. 
\label{fig:trivialDN}}
\end{minipage}
\end{center}
\end{figure}
%
%
\noindent
and thus $L_+$ is positive definite for sufficiently large $V_0$. 
This, in turn, implies instability of the sn$(x,k)$ solution for 
sufficiently large $V_0$, which corresponds to large amplitude 
solutions.  The sn$(x,k)$ solution goes quickly unstable in a 
similar fashion to the $\cn(x,k)$ solution (see Fig.~\ref{fig:trivialCN}).

%
%


\subsubsection{dn$(x,k)$}

From the previously established results, linear stability for the dn$(x,k)$
solutions follows immediately since $r(x)>0$, because dn$(x,k)$ has no zeros.
%
%
%
%
Thus in contrast to the cn$(x,k)$ and sn$(x,k)$ solutions,
the dn$(x,k)$ solutions given by
Eq.~(\ref{eqn:dn}) are linearly stable.  
Figure~\ref{fig:trivialDN} displays the evolution
of a dn$(x,k)$ solution over the time interval 
$t \in [0,80]$ for $V_0=-1.0$ and $k=0.5$.  
Although noise was added to the initial data, the solution 
shape persists and the solution
is stable, as predicted analytically. 
For this case, the wavenumber spectrum
is supported primarily by three modes:  the zero mode which determines the
offset, and two other modes which determine the oscillation frequency 
of the dn$(x,k)$ solution. Even with large perturbations, this
solution persists. This indicates that the offset of a solution is
important for its stability. This observation is reconfirmed for
other stable solutions below.   

\subsection{Trivial Phase:  Type B}

\subsubsection{cn$(x,k)$}

The Type B trivial phase solution is obtained for $a_1=\pm {k^2}/{4}$ and
corresponding amplitude $|r(x)|=({k}/{2})\sqrt{1+\cn(x,k)}$. The solution
$r(x)$ is not strictly positive. The operator $L_+$ is
\begin{equation}
  L_+ \!=\!-\demi\partial_x^2\!+\!\frac{k^2}{4}\!-\!\frac{1}{8} 
     \!+\!\frac{3}{8}k^2\sn^2 (x,k) \!+\!3a_1\cn (x,k). \!\!\!
\end{equation}
Thus we find that the situation 
is indeterminate.   

\begin{figure}[t]
\centerline{\psfig{figure=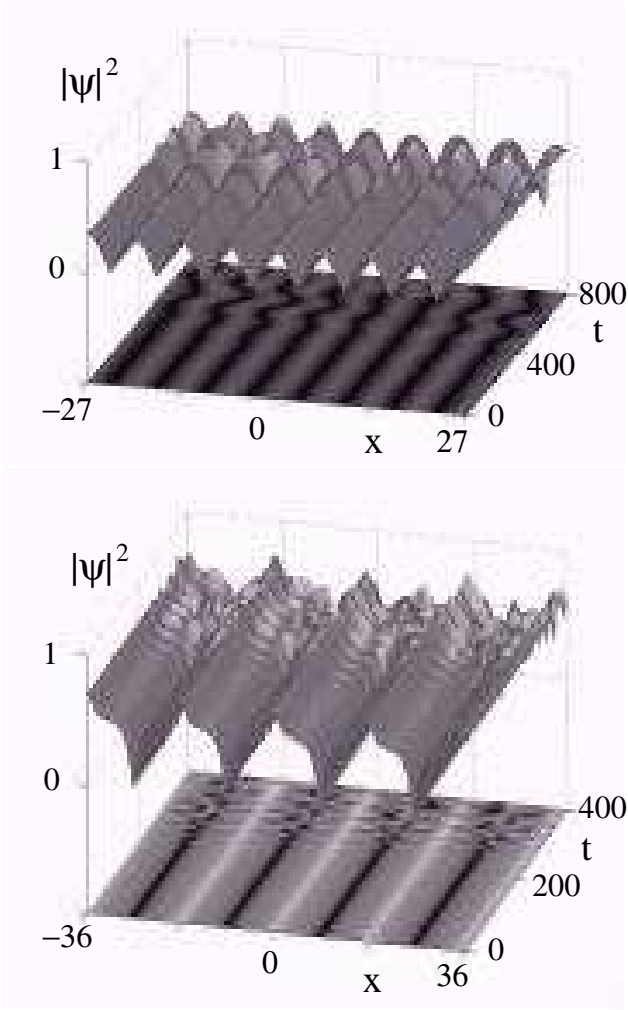,width=83mm,silent=}}
\begin{center}
\begin{minipage}{83mm}
\caption{Unstable evolution of the Type B cn$(x,k)$ solutions
given by Eq.~(\ref{eqn:trivphaseB1}) for $k=0.5$ (top panel) and
$k=0.999$ (bottom panel) for $a_1=k^2/4$ and $V_0=-3k^3/8$.
\label{fig:lin-CN}}
\end{minipage}
\end{center}
\end{figure}

\begin{figure}[t]
\centerline{\psfig{figure=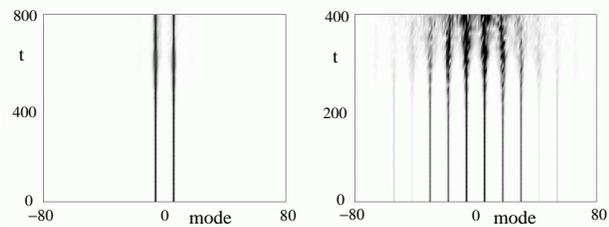,width=83mm,silent=}}
\begin{center}
\begin{minipage}{83mm}
\caption{Wavenumber spectrum evolution of a Type B cn$(x,k)$  
solution given by Eq.~(\ref{eqn:cn}) for $a_1=k^2/4$ and corresponding to 
$k=0.5$ (left panel) and $k=0.999$ (right panel) of Fig~\ref{fig:lin-CN}. 
The evolution shows that the unstable band of modes is generated
near wavenumber one and for $k=0.999$ near wavenumber one and its 
harmonics.
\label{fig:spec-lin}}
\end{minipage}
\end{center}
\end{figure}

Numerical simulations for the Type B cn$(x,k)$ solutions given by
Eq.~(\ref{eqn:trivphaseB1}) are illustrated in Fig.~\ref{fig:lin-CN}.  This
figure displays the evolution of the cn$(x,k)$ branch of solution for $k=0.5$
(top panel) and $k=0.999$ (bottom panel) over the time interval $t \in
[0,800]$ and $t\in [0,400]$ respectively for $V_0=-3k^2/8$.  For both $k=0.5$
and $k=0.999$ the solutions are unstable, but this instability manifests
itself only after several hundred time units.  Figure~\ref{fig:spec-lin} shows
the evolution of the wavenumber spectrum for both these cases.  For $k=0.5$,
the onset of instability occurs near wavenumber one as is the case of 
Type A
solutions.  After 800 time units, the wavenumbers have only just begun to
spread, causing the solution~to
%
%
%
%
%
%
%
%
\begin{figure}[t]
\centerline{\psfig{figure=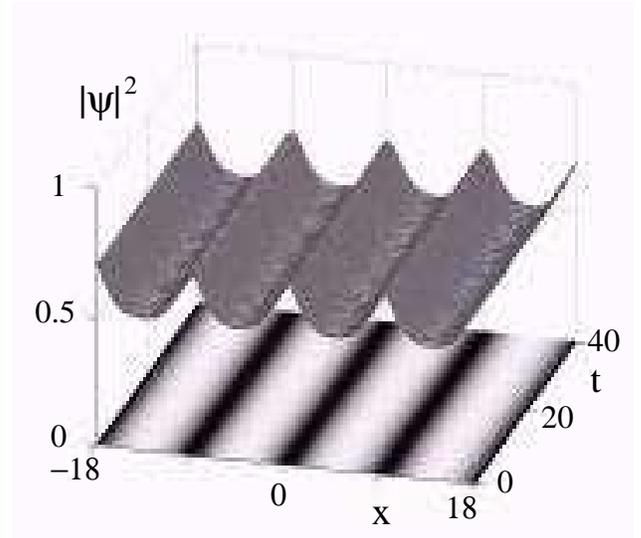,width=83mm,silent=}}
\begin{center}
\begin{minipage}{83mm}
\caption{Stable evolution of a Type B dn$(x,k)$ solution
given by Eq.~(\ref{eqn:trivphaseB2}) for $k=0.999$ and $a_2=1/4$.
\label{fig:lin-DN2}}
\end{minipage}
\end{center}
\end{figure}
%
%
\noindent
destabilize.  For $k=0.999$, the solution is
composed of a much larger number of wavenumbers which destabilize much more
quickly than the $k=0.5$ case.  Here the instability is generated near
wavenumber one and its harmonics.

\subsubsection{dn$(x,k)$}

The trivial phase dn$(x,k)$ solution requires $c=0$ which is achieved for
$a_2=\pm {1}/{4}$, $a_2=0$, or $a_2={\sqrt{1-k^2}}/{4}.$ 
Thus three distinct parameter regimes need to be considered.
The relevant operators in this case are 
\begin{subeqnarray} 
L_+&\!\!\!\!=\!\!\!\!&-\demi\partial_x^2\!-\!\frac{k^2\!+\!1}{8}\!+\!6a_2^2 
    \!\!+\!\!\frac{3k^2}{8}\sn^2 \! (x,\!k) \!+\!3a_2\dn (x,\!k), ~~~~~~~ \\
L_-&\!\!\!\!=\!\!\!\!&-\demi\partial_x^2\!-\!\frac{k^2\!+\!1}{8}\!-\!2a_2^2
             \!\!+\!\!\frac{3k^2}{8}\sn^2 \! (x,\!k) \!+\!a_2\dn (x,\!k).
\end{subeqnarray}
The case $a_2=1/4$ gives $L_-r(x)=0$ with $r(x)>0$.  Hence from the linear
stability criteria, these waves are stable for all values of $k$.  As with the
$a_2=1/4$ case, the regime where $a_2={\sqrt{1-k^2}}/{4}$ gives a solution
$r(x)$ which is strictly positive and is the ground state of $L_-$.
Thus stability follows for all values of $k$.  The last parameter regime, for
which $a_2=-1/4$, is indeterminate since both $\lambda_-$ and $\lambda_+$ are
negative and our linear stability analysis is inconclusive.

\begin{figure}[t]
\centerline{\psfig{figure=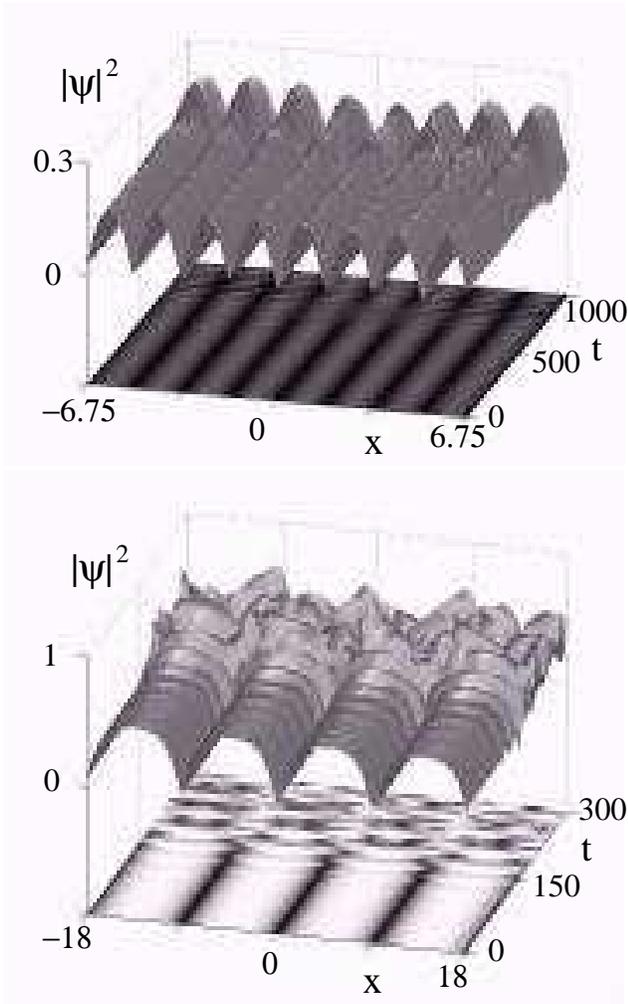,width=83mm,silent=}}
\begin{center}
\begin{minipage}{83mm}
\caption{Unstable evolution of a Type B dn$(x,k)$ solution
given by Eq.~(\ref{eqn:trivphaseB3}) for $k=0.5$ (top panel) and
$k=0.999$ (bottom panel) given $a_2=-1/4$.
In this case, there is no offset to stabilize the condensate.
\label{fig:lin-DN}}
\end{minipage}
\end{center}
\end{figure}

These analytic predictions are confirmed in
Figs.~\ref{fig:lin-DN2}--\ref{fig:lin-DN}. In Fig.~\ref{fig:lin-DN2} the
evolution of a dn$(x,k)$ solution is shown for $a_2=1/4$ and $k=0.999$.  As
predicted analytically, this parameter regime is stable for all $k$ values.
This simulation once again illustrates the importance of an offset for
stabilizing the condensate~\cite{prl}.  In contrast to this stable evolution,
the case $a_2=-1/4$ is unstable as illustrated in Fig.~\ref{fig:lin-DN}.  The
linear stability results in this case are indeterminate. ~However, the 
numerical~simulations 
%
%
%
%
%
\begin{figure}[t]
\centerline{\psfig{figure=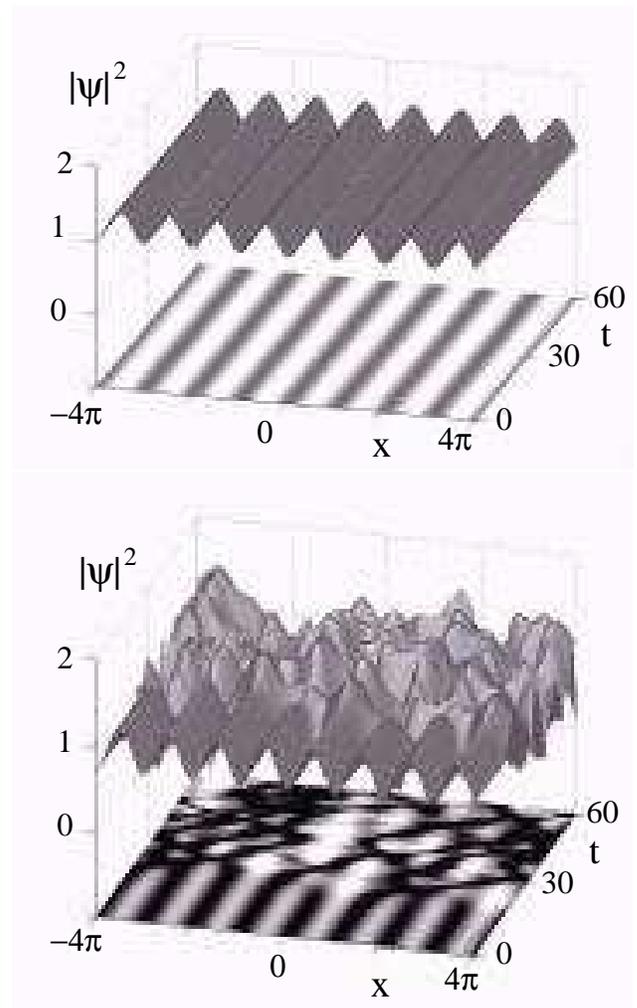,width=83mm,silent=}}
\begin{center}
\begin{minipage}{83mm}
\caption{Evolution of a nontrivial phase Type A solution with
$V_0=1.0$ and $B=1$ (top panel) and $B=1/2$ (bottom panel).
For $B$ sufficiently large, the offset provided is able
to stabilize the condensate whereas for $B$ below a critical
threshold the condensate destabilizes as shown for $B=1/2$.
\label{fig:nontrivialA}}
\end{minipage}
\end{center}
\end{figure}
%
%
\noindent
conclusively show the evolution to be unstable for all $k$ values.
For this case, the offset of the solution is insufficient to stabilize the
condensate.  We note that for small values of $k$, the onset of instability
occurs after a very long time.  Higher values of $k$ result in instabilities
on a much faster time scale.  Finally, we consider the parameter regime for
which $a_2={\sqrt{1-k^2}}/{4}$.  In this case, the analytic predictions once
again suggest stability for all $k$ values.  We do not illustrate this case
since it is qualitatively very similar to Fig.~\ref{fig:lin-DN}. However, in
contrast to the $a_2=1/4$ case, for values of $k$ close to one, there is a
negligible amount of offset, distinguishing this stable case from previous
ones.  For these values of $k$, the solution has a small amplitude compared to
the potential so that the behavior is essentially linear and stability is
achieved because the condensate is trapped in the wells of the potential, as
in ordinary quantum mechanics.


\subsection{Nontrivial Phase}

As stated at the beginning of Section \ref{sec:stability}, determining the
linear stability for nontrivial phase solutions is not amenable to analysis.
This leads us to consider the stability of nontrivial phase solutions using
numerical computations.

To begin, consider the trigonometric limit of the nontrivial phase solutions
of Type A.  These solutions are given by Eqs.~(\ref{eqn:trig}) and
(\ref{eqn:genphase}).  Figure~\ref{fig:nontrivialA} depicts the evolution of a
pair of initial conditions with $V_0=1.0$ and for which $B=1$ (top panel) and
$B=1/2$ (bottom panel).  Since $B$ determines the offset of the condensate,
these numerical results show directly the importance of this offset for
stability.  In contrast, if the offset is too small, it is unable to stabilize
the condensate.


For Type B solutions, qualitatively nothing changes from the
dynamics illustrated for the trivial phase case.  In particular, numerical
simulations can be performed using exact solutions which are constructed
subject to the phase quantization condition given by either
Eq.~(\ref{eqn:genquant1}) or (\ref{eqn:genquant2}).  A numerical shooting
method is used to find appropriate values of $a_2$ for which a
phase--quantized, periodic solution exists.  Once this is achieved, numerical
simulations can easily be performed. Note that any integer value $p$ is allowed
as input for the phase quantization conditions, provided solutions exist for
the parameter values. It turns out this imposes a lower bound on value of $p$.
In the simulations, the actual value of $p$ does not affect the stability of
the solution.  Increasing the phase--quantization integer $p$ leads to a
solution with a steeper phase profile, suggesting a more unstable situation.
However, this phase effect is balanced by an increased offset $a_2$ of the
amplitude.  Qualitatively, the dynamics are as depicted in
Figs.~\ref{fig:lin-DN2}--\ref{fig:lin-DN}.  Thus the nontrivial phase solutions
of Type B are stable for $a_2>1/4$ and for $0<a_2<\sqrt{1-k^2}/4$, whereas the
nontrivial phase solution is unstable for $a_2<-1/4$.







\section{Summary and Conclusions}

We considered the repulsive nonlinear Schr\"odinger 
equation with an elliptic function potential as a 
model for a trapped, quasi-one-dimensional Bose-Einstein 
condensate.   Two new families of periodic solutions of 
this equation were found and their stability was investigated both
analytically and numerically.  Using analytical results for 
trivial phase solutions, we showed that solutions with sufficient offset
are linearly stable.  
Moreover, all such stable solutions are 
deformations of the ground state of the linear Schr\"odinger equation.
This is confirmed with extensive numerical
simulations on the governing nonlinear equation.  
Likewise, nontrivial phase
solutions are stable if their density is sufficiently offset.
%
%
%
Since we are modeling a Bose-Einstein
condensate trapped in a standing light wave, our results imply that
a large number of condensed atoms is sufficient 
to form a stable, periodic condensate.  Physically, this implies
stability of states near the Thomas--Fermi limit.

To quantify this phenomena, we consider the $k=0$ limit and note 
that from Eqs.~(\ref{eqn:NLS}) and~(\ref{eqn:trig}), the number
of particles per well $n$ is given by
$n=(\int_0^\pi |\psi(x,t)|^2 dx)/\pi=V_0/2 + B$.  
In the context of the BEC, and for a fixed atomic coupling strength,
this means a large number of condensed atoms    
per well $n$ is sufficient to provide an offset on the order of 
the potential strength.  This ensures stabilization of the condensate.
Alternatively, a condensate with a large enough number of atoms can be 
interpreted as a developed condensate for which the nonlinearity 
acts as a stabilizing mechanism.

{\bf Acknowledgments:} We benefited greatly from discussions with Ricardo
Carretero-Gonz\'alez and William Reinhardt.   The work of J. Bronski, 
L. D. Carr, B. Deconinck, and J. N. Kutz
was supported by National Science Foundation Grants DMS--9972869, CHE97--32919,
DMS--0071568, and DMS--9802920 respectively. K. Promislow acknowledges support
from NSERC--611255



\begin{thebibliography}{10}
\bibitem[*]{byline} to whom correspondence should be addressed
\end{thebibliography}

\end{multicols}
\end{document}